\documentclass[%
 reprint,
 amsmath,amssymb,
 aps,
prb,
]{revtex4-2}

\usepackage{graphicx}
\usepackage{dcolumn}
\usepackage{bm}

\begin{document}

\preprint{APS/123-QED}

\title{Scaling regimes of the one-dimensional phase turbulence in the deterministic complex Ginzburg-Landau equation}

\author{Francesco Vercesi $^1$, Susie Poirier $^1$, Anna Minguzzi$^1$, L\'eonie Canet$^{1,2}$ }
\affiliation{
$^1$ Université Grenoble Alpes, CNRS, LPMMC, 38000 Grenoble, France, \\
$^2$ Institut Universitaire de France, 5 rue Descartes, 75005 Paris}


\begin{abstract}
We consider the one-dimensional deterministic complex Ginzburg-Landau equation in the regime of phase turbulence,
where the order parameter displays a defect-free chaotic phase dynamics mapping to the Kuramoto-Sivashinsky equation, characterized by negative viscosity and a modulational instability at linear level.
In this regime, the dynamical behavior of the large wavelength modes is captured by the Kardar-Parisi-Zhang (KPZ) universality class, determining their universal scaling and their statistical properties. These modes exhibit the characteristic KPZ sub-diffusive scaling with the  dynamical critical exponent $z=3/2$. We present numerical evidence of the existence of an additional scale-invariant regime, with the dynamical exponent $z=1$, emerging at scales which are intermediate between the microscopic ones, intrinsic to the modulational instability, and the macroscopic ones. We argue that this new scaling regime belongs to the universality class corresponding to the inviscid limit of the KPZ equation.

\end{abstract}

\maketitle

\section{Introduction}

The complex Ginzburg-Landau equation (CGLE) is a prototype model for the effective dynamics of spatially extended systems out of equilibrium, ranging from hydrodynamical instabilities \cite{Manneville1990}, pattern formation \cite{Cross2009}, chemical turbulence \cite{Aranson2002}, to driven-dissipative bosonic condensates \cite{Carusotto2013}.
Its success in providing a reliable qualitative description of a vast variety of phenomena in terms of a few parameters has earned the CGLE a conspicuous interest \cite{Aranson2002}.  Important efforts have  been devoted  to characterizing the rich phase diagram of the CGLE, both in one and in higher spatial dimensions \cite{Sakaguchi1990, Shraiman1992,Chate1994, Chat1996, Aranson2002}.
In one dimension, the CGLE can yield chaotic, non-chaotic and intermittent dynamics. 
We focus in this paper on the weakly turbulent regime, also known as phase turbulence, characterized by spatio-temporal chaos in absence of topological defects
\cite{Sakaguchi1990, Egolf1995, Montagne1996, Montagne1997, Torcini1997, Baalen2004}.  In this regime, the amplitude of the order parameter weakly fluctuates  around a finite steady value. Its dynamics can be integrated out resulting in a mapping of the CGLE to the Kuramoto-Sivashinsky (KS) equation \cite{Kuramoto1978, Sivashinsky1977} governing the effective dynamics of the phase.  The KS equation is a 
deterministic nonlinear model for fluctuating interfaces, which exhibits nontrivial behavior: indeed, this equation features a negative viscosity, which yields an intrinsic modulational instability of the linearized equation, saturated at 
nonlinear level. This induces a chaotic dynamics when the size of the system
is large with respect to the typical scale of the instability pattern \cite{Hyman1986, Papageorgiou1991}. 
This dynamics exhibits a steady state, whose 
 essential statistical features are captured by the celebrated Kardar-Parisi-Zhang (KPZ) equation \cite{Kardar1986}.

Originally introduced for modeling the random growth of non-equilibrium interfaces \cite{Kardar1986}, the KPZ equation is a stochastic nonlinear partial differential equation which has become a paradigm of nonequilibrium criticality, encompassing a wide collection of systems counting, besides driven rough interfaces, randomly stirred viscous fluids \cite{Burgers1948}, directed polymers in random media \cite{Krug1992}, the coherence of driven-dissipative bosonic condensates \cite{Ji2015, He2015, Fontaine2022}, quantum spin chains \cite{Moca2023}, strongly correlated bosons \cite{Fujimoto2020}, and many more \cite{Takeuchi2018}.
In one dimension, the KPZ equation yields a universal critical regime characterized by a  sub-diffusive scaling behavior with the exact dynamical critical exponent $z=3/2$ \cite{Kardar1986}, as well as by precisely known non-Gaussian statistics \cite{Prhofer2004, Takeuchi2018}.
In the limit of vanishing nonlinearity, the KPZ equation reduces to the Edwards-Wilkinson (EW) equation \cite{Edwards82}, which leads to a diffusive scaling with $z=2$ and Gaussian statistics. Recently, a new scaling regime, characterized by a dynamical exponent $z=1$,
 has been unveiled in different systems belonging to the KPZ universality class \cite{Majda2000, Brachet2022, Cuerno2022, Fujimoto2020, Fontaine2023}. This new regime emerges in the limit of vanishing surface tension for the KPZ equation, or equivalently in the limit of vanishing viscosity for the stochastic Burgers equation,  which governs the dynamics of the velocity of the interface \cite{Majda2000, Brachet2022, Cuerno2022, Fujimoto2020}.
The functional renormalization group analysis carried out in Ref. \cite{Fontaine2023} has  shown that this new scaling regime is controlled by a fixed point of the KPZ equation which had not yet been identified, and  which corresponds to its inviscid limit. This fixed point was termed ``inviscid Burgers'' (IB) fixed point and it describes a new universality class, featuring in particular the $z=1$ dynamical exponent. Its associated universal scaling function was determined  using the functional renormalization group in Ref.~\cite{Fontaine2023}.

The fact that the large-scale behavior of the deterministic KS equation belongs to the KPZ universality class was early conjectured \cite{Hyman1986}, and supported by some indications in  numerical simulations \cite{Sneppen1992} and perturbative renormalization group analysis of the noisy KS equation \cite{Cuerno1995, Ueno2005}. Yet, the clear signature of the KPZ universal scaling in this system has long eluded  numerical observation due to the very large system size and run time needed. The  scaling regime found in early simulations was rather compatible with the EW one.
 A definitive evidence of KPZ scaling has only been recently provided by the massive numerical simulations performed by the authors of Ref. \cite{Roy2020}, which allowed them to quantitatively determine the scaling exponents and the statistical properties of the one-dimensional deterministic KS equation, and show that they correspond to  the KPZ ones.

In this paper, we present a numerical study of the statistical properties of the deterministic CGLE in the phase turbulence regime, and show that, besides the  KPZ universal scaling, also the Inviscid Burgers one arises. We  provide arguments to explain the systematic appearance of the IB universality in the CGLE, and in the Kuramoto-Sivashinsky equation.
 In details, we simulate the deterministic CGLE,  starting from random initial conditions, and we compute the spatio-temporal correlations of the phase by averaging over independent realizations. We first show that the  scaling behavior of the long wavelength modes is the EW one ($z=2$), as expected for to the system size considered.
 We then focus on intermediate scales and show that they exhibit a different scaling, with $z=1$, corresponding to the IB universality class. We compute the associated scaling function. Finally, as proposed for the KS equation in Ref. \cite{Ueno2005}, we introduce a stochastic white noise in the CGLE and 
 identify a window of parameters in which the KPZ universality 
 emerges, despite the small system size and the 
 finite probability of noise-activated defect formation
  \cite{He2017, Vercesi2023}.

\section{\label{sec:Model} The deterministic complex Ginzburg-Landau equation}

\subsection{\label{subsec:PT} Regime of phase turbulence}
We consider the complex Ginzburg Landau equation defined by:
\begin{equation}
    i\partial_t \psi = i\psi + (c_2 - i)|\psi|^2 \psi - (c_1 - i) \partial_x^2 \psi
    \label{eq:CGLE_c1c2}
\end{equation}
where $\psi$ is the complex order parameter and $c_1$, $c_2$ are dimensionless real coefficients. 
The homogeneous solution $\psi_0 = 1 \times \mathrm{e}^{-i c_2 t}$ is linearly unstable when $1 + c_1c_2 < 0$.
Under this condition, a modulational instability, which is known as the Benjamin-Feir (BF) instability \cite{Benjamin1967, Aranson2002}, triggers a turbulent behavior which, depending on the values of $c_1$ and $c_2$, is either characterized by the presence of topological defects, where the amplitude $|\psi|$ goes to zero, or by defect-free phase modulations \cite{Aranson2002}. The two regimes are usually labeled defect and phase turbulence, respectively.
The latter arises if $c_1, c_2$ are chosen close to the BF instability line $1 + c_1c_2 = 0$ \cite{Baalen2004}, in which case the amplitude slightly fluctuates around 1, while the phase dynamics is mapped to the Kuramoto-Sivashinsky equation \cite{Kuramoto1978, Sivashinsky1977}:
\begin{equation}
    \partial_t \theta = \left(\nu \partial_x^2  + \eta \partial_x^4 \right) \theta + \frac{\lambda}{2} (\partial_x \theta)^2
    \label{eq:KSE_phase}
\end{equation}
with $\nu = 1+c_1c_2$, $\eta = -c_1^2 / 2$, $\lambda = 2(c_2 - c_1)$. 
We report the derivation of this phase equation in Appendix \ref{app:mapping_phase}.
One readily notices that the instable regime of the CGLE corresponds to a negative value of the viscosity $\nu$ in the phase equation (\ref{eq:KSE_phase}). The instability  primarily concerns the low momentum modes, since the linear dispersion $-\nu k^2 + \eta k^4$ is positive for $0 \leq k < k_0$ with $k_0 = \sqrt{\nu / \eta}$. In the following, we focus on the regime of phase turbulence, by appropriately choosing $c_1$ and $c_2$ close to the BF instability line.

\subsection{Large scale behavior}

The statistical properties of the large wavelength fluctuations of the KS phase  are expected to belong to the KPZ universality class \cite{Hyman1986}, which means that their effective macroscopic dynamics can be described by the KPZ equation \cite{Kardar1986}:
\begin{equation}
    \partial_t \theta = \nu_{\rm eff} \partial_x^2 \theta + \frac{\lambda_{\rm eff}}{2} (\partial_x \theta)^2 + \xi(x, t)
    \label{eq:KPZ_eff}
\end{equation}
where $\nu_{\rm eff}>0$ and $\xi(x,t)$ is a white noise with $\langle \xi(x, t) \xi(x',t') \rangle = 2 D_{\rm eff}\delta(x-x')\delta(t-t')$.
However, it was shown that a sufficiently large system size is required for KPZ universality to emerge, while smaller systems were observed  to display EW scaling \cite{Roy2020}.

We emphasize that if one considers from the start a noisy version of the CGLE (\ref{eq:CGLE_c1c2}) and chooses $c_1$, $c_2$ within the stable region (implying $\nu>0$ in (\ref{eq:KSE_phase})), the term proportional to $\partial_x^4$ becomes subdominant and  the phase dynamics simply inherits the stochastic nature of the CGLE. In this case, one directly   obtains the KPZ equation  \cite{He2015, Ji2015}.
This is the case for driven-dissipative bosonic condensates described by the generalized stochastic Gross-Pitaevskii equation, in which the large-scale coherence, controlled by phase fluctuations, was shown to exhibit the KPZ scaling \cite{Ji2015, He2015, Fontaine2022} and non-Gaussian statistics \cite{Squizzato2018, Deligiannis2021} even in small systems.

Conversely, in the deterministic case and in the unstable regime where $\nu<0$, the phase dynamics maps to the deterministic KS equation, and the parameters $\nu_{\rm eff}$ and $D_{\rm eff}$ of the effective KPZ equation are not primarily determined, but are generated by the chaotic dynamics.  The perturbative renormalization group analysis of Ref. \cite{Ueno2005} showed the emergence of a positive macroscopic viscosity from the noisy microscopic KS equation.
Indeed, they find that while coarse-graining by including the fluctuations from small to large scales,  the effective viscosity changes sign, from the microscopic value $\nu<0$  to an effective  value $\nu_{\rm eff}>0$ in the macroscopic limit. We emphasize that in this process, the viscosity crosses zero at some intermediate scale: 
this is the very origin of the appearance of the IB regime at intermediate scales \cite{ks_kpz_prep2024}.  Note that the Galilean invariance of the KS equation imposes $\lambda_{\rm eff} = \lambda$ at all scales, although in practice the value of $\lambda_{\rm eff}$ may be renormalized by effect of the space discretization \cite{Sneppen1992} or, as in the present case, of the higher-order nonlinearities neglected when mapping the CGLE to the KS equation (see Appendix \ref{app:mapping_phase}).

In order to characterize the statistical properties of the phase dynamics, we analyze both the transient regime and the steady state. In the transient regime, i.e. during the kinetic roughening, the dynamical scaling behavior of the interface $\theta(x, t)$ is captured by  the structure function, defined as
\begin{equation}
    S(k, t) = \langle \theta(k, t) \theta(-k, t)  \rangle
    \label{eq:S}
\end{equation}
where in the deterministic case $\langle . \rangle$ denotes the average over independent trajectories with randomly chosen initial conditions, while for the stochastic equation, in Sec. \ref{subsec:noise_KPZ}, the average is over independent noise realizations.
The choice \eqref{eq:S} is particularly suited when different scaling behaviors coexist at different scales $k$ \cite{Ramasco2000}.
The Family-Vicsek scaling Ansatz \cite{Family1985} for $S(k, t)$ reads:
\begin{equation}
    S(k, t)  = k^{-(2\chi + 1)} s(k^z t)
    \label{eq:def_Skt}
\end{equation}
where $\chi$ is the roughness critical exponent, $z$ is the dynamical critical exponent, and $s(y)$ is a universal scaling function with the properties $s(y\rightarrow 0)\sim y^{2\chi + 1}$ and $s(y\rightarrow \infty) \rightarrow s_0$.
For each mode $k$, the structure function converges to the stationary average occupation when the time $t_{ss}(k) \sim k^{-z}$ has elapsed.

In the steady state, we focus on the velocity field $u=\partial_x \theta$ and compute the temporal correlations of its Fourier modes $C(k, t) = \langle u(k, t + t_0) u^*(k, t_0) \rangle$, with $t_0 > t_{ss}(k)$ for all the modes $k$ considered.
The correlations exhibit the scaling behavior
\begin{equation}
    C(k, t) = C(k, 0) f\left( k^{z}t \right)
    \label{eq:def_Ckt}
\end{equation}
where $f(y)$ is a universal scaling function and the exponents $\chi$, $z$ are the same as in Eq.~(\ref{eq:def_Skt}). Note that $t$ in Eq.~(\ref{eq:def_Skt}) is the absolute time whereas  it  denotes in Eq.~(\ref{eq:def_Ckt}) the time delay.

\subsection{KPZ fixed points}
The KPZ equation~\eqref{eq:KPZ_eff} can be rescaled to obtain a single relevant parameter $g  = \lambda ^2 D  / \nu ^3$.
Depending on the value of $g $, there are three possible scaling regimes, controlled by the corresponding fixed points:
\begin{enumerate} 
    \item[i)] the KPZ regime ($g $ finite): $\chi=1/2$, $z=3/2$ and $f = f^{\rm KPZ}$ given by the universal KPZ scaling function, exactly calculated in Ref. \cite{Prhofer2004}
    \item[ii)] the EW regime ($g  = 0$): $\chi=1/2$, $z=2$ and $f = f^{\rm EW}$ the universal scaling function of the linear theory, given by
    \begin{equation}
        f^{\rm EW}(y) = \frac{ D }{\nu }e^{-\sqrt{\nu}y^2}
        \label{eq:fy_EW}
    \end{equation}
    
    \item[iii)]   the IB regime ($g  = \infty$): $\chi=1/2$, $z=1$ and $f = f^{\rm IB}$ computed via the functional renormalization group in Ref. \cite{Fontaine2023}. The short-time asymptotic behavior of $f^{\rm IB}$ was shown to endow a simple Gaussian form \cite{Majda2000, Brachet2022, Fontaine2023}:
    \begin{equation}
        f^{\rm IB}(y \ll y_0) \sim \, e^{-ay^2}\, .
        \label{eq:fy_IB_asympt}
    \end{equation}
\end{enumerate}

In one dimension,  the only attractive fixed point in the infrared (at large distance) for any finite $g$ is the KPZ fixed point. However, in a finite-size system, the other two EW and IB fixed points, although repulsive in the infrared, can influence the scaling properties of the system when $g$ is respectively very small or very large, especially for the intermediate modes \cite{Brachet2022, Fontaine2023}.

Let us  emphasize that  in one dimension all the three regimes of the KPZ equation share a common value $\chi=1/2$ for the roughness exponent. This implies that the stationary state is characterized by a flat energy spectrum
$C(k, 0) = \langle |u(k)|^2 \rangle = k^2 S(t>t_{ss}, k) = s_0 = \frac{2D}{\nu}$. 
The equipartition of energy is a consequence of the ``accidental'' time-reversal symmetry  of the KPZ equation in one dimension, whereby the spatial properties of the stationary interface coincides with the equilibrium one (i.e. the Brownian interface of the diffusive EW case) \cite{Takeuchi2018}.
The associated fluctuation-dissipation relation \cite{Canet2011} constrains the effective macroscopic viscosity and noise strength to conserve the ratio they have at the microscopic level. We underline that the ratio $\nu_{\rm eff}/D_{\rm eff}$ is not defined for the deterministic KS equation, for which the time-reversal symmetry is an emergent property at large scales~\cite{Ueno2005}.

More subtle is the inviscid limit of the KPZ (Burgers) equation. Indeed,  the typical solutions of the inviscid Burgers equation in an infinite system generate dissipative shocks in a finite time \cite{Burgers1948, burgulence}. However, in the presence of an ultraviolet cutoff and if the time evolution is  energy-conserving,  the system evolves instead to a thermalized state  with the equilibrium static exponent $\chi=1/2$, and a dynamical exponent $z=1$ \cite{Majda2000, Brachet2022, Fontaine2023}.
This implies  that the inviscid limit $\nu \rightarrow 0$ is approached by preserving a constant finite ratio $\nu / D$, thus coinciding with the deterministic limit.
As a final remark, we mention that the authors of Ref. \cite{Cuerno2022} characterized the scaling behavior of an inviscid and noisy version of the KPZ equation, which breaks the time-reversal symmetry. In this case, they report an anomalous kinetic roughening behavior with different local and global roughness exponents ($\chi_{\rm loc} = 1 \neq \chi = 1/2$), while conserving the 
dynamical exponent $z=1$.
To fully understand the connection between this regime and the thermalized one of Ref. \cite{Majda2000, Brachet2022, Fontaine2023} is an exciting, although non-trivial task.

\section{\label{sec:Results} Results}

We have performed  numerical simulations of the CGLE~(\ref{eq:CGLE_c1c2}) with random initial conditions $\psi(t=0, x) = 1 + \sigma(x)$, where $\sigma(x) \in \mathbb{C}$ is a random complex number drawn independently for every $x$ from a Gaussian distribution of zero mean and variance 0.01 \footnote{The winding number, defined as $w = \frac{1}{2\pi} \int_0^L \partial_x\theta dx$, is thus always zero in our initial conditions. In the phase turbulence regime $w$ is conserved. We exclude the \textit{drifted phase turbulence} scenario of Ref. \cite{Montagne1996}, obtained by setting $w\neq0$ at $t=0$, in which case the KS dynamics develops on top of a plane wave of wavenumber $k = 2\pi w / L$.}. 
We consider the equation in its dimensionless form \eqref{eq:CGLE_c1c2}, with a space discretization chosen as $dx=1.0$ or $dx=1.5$ -- the change  corresponding to a slightly modified weight of the kinetic term with respect to the others in the CGLE.
We study the statistical properties of the interface defined as the unwrapped phase $\theta(x, t) = \mathrm{Arg}(\psi(x, t)) + 2\pi j$, where $j$ is an integer chosen in order for the temporal evolution to be smooth in time, i.e. $\theta(x, t + \delta t) - \theta(x, t) \in (-\pi, \pi]$. We first present the results for the large-scale behavior, which exhibits the EW scaling,  then discuss the behavior of the intermediate scales where we find the IB scaling, and finally show that, adding a small noise, the KPZ scaling emerges at the large scales, replacing the EW regime.
\begin{figure*}[t]
    \centering
    \includegraphics[width=.8\linewidth]{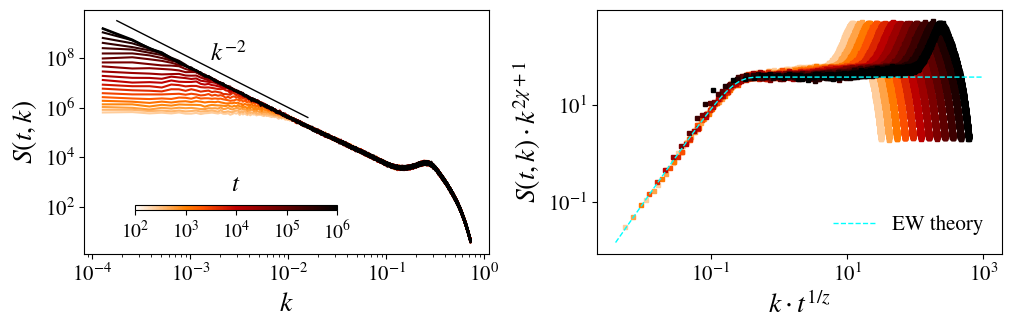}
    \caption{Scaling behavior of the structure function compared with the EW theory (\ref{eq:fy_EW}), with $\chi=1/2$, $z=2$. The plotted window is $2\pi/L < k< 2 k_0$. The parameters are: $c_1=3.5$, $c_2=-0.6$ ($k_0=0.36$), $L=49152$, $dx=1.5$, $N_{\rm sim}=1024$}
    \label{fig:scaling_EW}
\end{figure*}

\subsection{\label{subsec:EW} Kinetic roughening behavior}

The structure function $S(t,k)$ in the transient roughening regime is displayed in Fig.~\ref{fig:scaling_EW}. During the early evolution, i.e. for $t \lesssim 500$ - not represented, the population of unstable low momentum modes $k<k_0$ is transferred to the dissipative sector $k>k_0$ via nonlinear coupling, until a typical cellular structure arises, whose wavenumber  corresponds to the local maximum
observed in $S(t,k)$.
Thereafter, the roughening process progressively fills the low momentum sector. The structure function is found to endow the scaling Ansatz (\ref{eq:def_Skt}), and the collapse obtained with the EW exponents $z=2$, $\chi=1/2$, shown in Fig.~\ref{fig:scaling_EW}, is excellent. The scaling function coincides with the EW one given by $s^{\rm EW}(y) = \frac{2 D_{\rm eff} }{\nu_{\rm eff} } \left(1 - \, e^{-2 \sqrt{\nu_{\rm eff} } y^2}\right)$ with the argument $y = kt^{1/2}$. We find $\nu_{\rm eff} \approx 12 $ and $D_{\rm eff} \approx 4.7 \times 10^{-3}$. These numbers, together with the effective non-linearity $\lambda_{\rm eff} \approx \lambda = 2(c_1 - c_2)$, give the KPZ coupling $g_{\rm eff} \approx 1.7 \times 10^{-4} $. Following Refs. \cite{Sneppen1992, Grinstein1996}, we can estimate the threshold system size and crossover time needed for the effective KS dynamics to exhibit the KPZ behavior, for which we obtain respectively $L_c \approx 6 \times 10^5 $, $t_c = 6 \times 10^8$, thus confirming \textit{a posteriori} the 
expectation of the EW scaling emerging for our system size.

\begin{figure}[h]
    \centering
    \includegraphics[width=.6\linewidth]{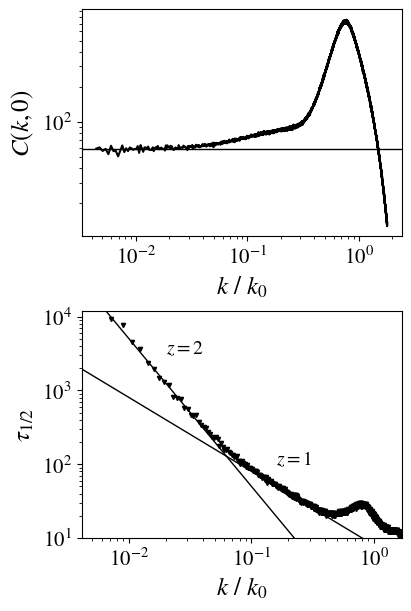}
    \caption{{\it Top panel}: energy spectrum $C(0,k)=\langle |u(k)|^2 \rangle$. {\it Bottom panel}: correlation time $\tau_{1/2}$, defined as $C(\tau_{1/2}, k) = \frac{1}{2} C(0, k)$. The parameters are: $c_1=2.1$, $c_2=-0.8$ ($k_0=0.434$), $L=8192$, $dx=1.0$, $N_{\rm sim}=1024$. The solid lines represent the $\sim k^{-z}$ behaviors.}
    \label{fig:spectrum_tau}
\end{figure}

\subsection{\label{subsec:IB} Inviscid Burgers regime}

\begin{figure*}[t]
    \centering
    \includegraphics[width=.8\linewidth]{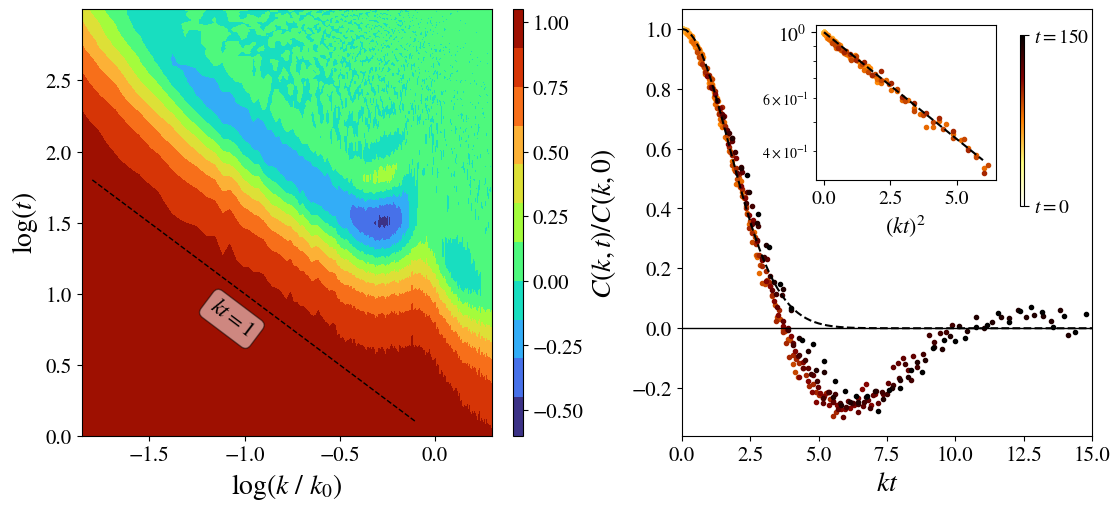}
    \caption{Scaling behavior of the correlations $C(k, t)$ in the IB regime. {\it Left panel}: correlation decay $C(t, k)/C(0, k)$. {\it Right panel}: collapse onto the scaling function $f(y=kt^{1/z})$ with $z=1$. The plotted window is $2\pi/L < k< 2 k_0$. The dashed line shows the fitted asymptotic Gaussian behavior of $f^{\rm IB}$ at short times, also plotted in logarithmic scale in the inset.
  The   parameters are: $c_1=2.7$, $c_2=-0.8$ ($k_0=0.44$), $L=2048$, $dx=1.0$, $N_{\rm sim}=1024$.}
    \label{fig:scaling_IB}
\end{figure*}

We now focus on the intermediate-scale modes, for which the spectrum becomes stationary at  short times, typically for $t \gtrsim 500$. 
We analyze the correlations $C(t, k)$ in the stationary state. The energy spectrum, given by $C(k, 0)$, is displayed in Fig.~\ref{fig:spectrum_tau}. It exhibits a plateau at small $k$ resulting from equipartition of energy.
 In order to identify the potential scaling regimes, we first compute the scale-dependent correlation time $\tau_{1/2}(k)$ defined for each mode $k$ from $C(t=\tau_{1/2}, k)=\frac{1}{2}C(t=0, k)$. It is expected to behave as $\tau_{1/2}\sim k^z$ in a scaling regime. The result, displayed in Fig.~\ref{fig:spectrum_tau}, clearly shows two distinct scaling behaviors: for small $k$, we retrieve the diffusive EW regime with $z=2$, while at intermediate $k$ we observe a different power-law dependence which is $\tau_{1/2} \sim k^{-1}$. To further characterize this scaling regime, we select the modes $k$ within the $z=1$ scaling region, and compute the full correlation $C(k, t)$, which is shown in Fig.~\ref{fig:scaling_IB}. The level lines coincide with constant $y=kt$, which indicates that $z=1$ in the scaling Ansatz (\ref{eq:def_Ckt}).
 We indeed obtain a remarkable collapse of the spatio-temporal data when plotted in the scaling variable $kt$, as shown in Fig.~\ref{fig:scaling_IB}. Furthermore, the short-time behavior of $f(y)$ is found in excellent agreement with the analytical prediction  in the IB regime, Eq.~(\ref{eq:fy_IB_asympt}), obtained by the FRG calculation  \cite{Fontaine2023}. The full shape of  $f(y)$ is also qualitatively compatible with the FRG result for $f^{\rm IB}$  and with its estimate in the numerical simulations of the Galerkin-truncated inviscid Burgers equation \cite{Brachet2022, Majda2000}, which both exhibit  a negative dip at a finite value of $y$ after the initial Gaussian decay. 
 However, the depth and position of the negative  dip we observe here do not  quantitatively compare with the values found in these works. This discrepancy is likely to originate in another essential difference of our work with the ``ideal'' IB regime of Refs. \cite{Brachet2022, Majda2000,Fontaine2023}. At the intermediate scales considered here,  the roughness exponent is slightly lower than $\chi = 1/2$, as can be observed in the energy spectrum of  Fig.~\ref{fig:spectrum_tau}, where a residual slope replaces the plateau for these intermediate modes.
The extracted value for $\chi$ has been found to depend on the parameters, ranging from $0.28$ to $0.43$ when $c_1$, $c_2$ take values within the square (2.70, -0.80), (2.10, -0.80), (2.10, -0.60), (2.70, -0.60). 
This signals that the equipartition of energy is not perfectly established for the modes that exhibit the IB scaling $z=1$.
We remind that in order for the equipartition to settle while approaching the inviscid limit of the stochastic Burgers equation, it is crucial that the noise obeys the fluctuation-dissipation relation \cite{Brachet2022}. In our case,  the random noise and the positive viscosity are effective properties of the small $k$ modes, emerging from the underlying chaotic dynamics of the CGLE. As a consequence, it is not guaranteed that they fulfill the time-reversal symmetry. We can thus reasonably expect a quantitative difference in the shape of the scaling function $f^{\rm IB}$.
A more detailed characterization of the IB regime emerging within the KS equation can be obtained 
 by means of functional renormalization group \cite{ks_kpz_prep2024}.
 Nevertheless, the results presented here suggest that the value of the dynamical exponent $z=1$ and the qualitative shape of $f^{\rm IB}$ are robust against a weak loss of equipartition.
 The more general analyzis of the inviscid regime for different types of noise, breaking the time-reversal symmetry would be interesting, but is left for future work.

\begin{figure}[h]
    \centering
    \includegraphics[width=.8\linewidth]{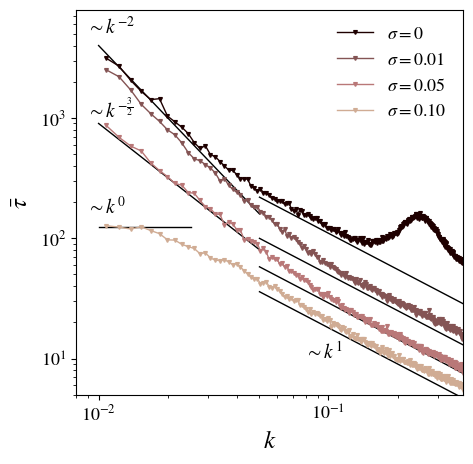}
    \caption{Correlation time $\bar{\tau}$ averaged over 10 values of $\alpha \in [0.3, 0.6]$ where $\alpha = C(k, \tau_{\alpha})/C(k,0)$. The parameters are: $c_1=1.9$, $c_2=-0.70$ ($k_0=0.35$), $L=4096$, $dx=1.0$, $N_{\rm sim}=1024$.
    The solid lines are guidelines  to identify the different scaling behaviors.}
    \label{fig:kpz_tau_noise_crossover}
\end{figure}

\subsection{\label{subsec:noise_KPZ} Adding noise: crossover to KPZ}

\begin{figure*}[t]
    \centering
    \includegraphics[width=.8\linewidth]{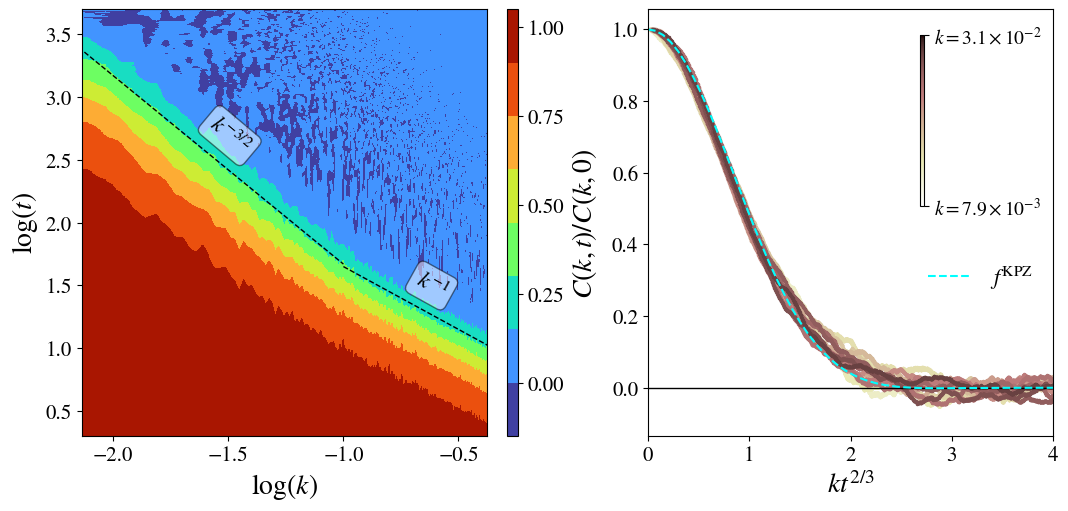}
    \caption{Scaling behavior of the correlations $C(k, t)$ in the KPZ regime. {\it Left panel}: correlation decay $C(t, k)/C(0, k)$. {\it Right panel}: collapse onto the scaling function $f(y=kt^{1/z})$ with $z=3/2$. The plotted window is $2\pi/L < k< 2 k_0$. The dashed line shows the exact KPZ scaling function $f^{\rm KPZ}$ from \cite{Prhofer2004}.
    The parameters are: $c_1=1.9$, $c_2=-0.7$ ($k_0=0.35$), $\sigma = 0.05$, $L=4096$, $dx=1.0$, $N_{\rm sim}=1024$.}
    \label{fig:scaling_KPZ}
\end{figure*}
In this section, we discuss the effect of adding a stochastic noise of small amplitude to the CGLE on the large-scale scaling behavior of the phase turbulence. We thus consider the equation
\begin{equation}
    i\partial_t \psi = i\psi + (c_2 - i)|\psi|^2 \psi - (c_1 - i) \partial_x^2 \psi + \sqrt{\sigma}\xi
    \label{eq:noisy_CGLE}
\end{equation}
where $\xi(x, t)$ is a complex white noise with $\langle \xi(x, t) \rangle = 0$ and $\langle \xi(x, t) \xi*(x', t') \rangle = 2 \delta(x-x')\delta(t-t')$.
We  replace the statistical average over trajectories starting  from different initial conditions with an average over independent realizations of the noise.

For the KS equation, the stochastic formulation yields considerable  advantages. At a theoretical level, it allows one to cast the problem into a field theory and study it by means of dynamical renormalization group \cite{Cuerno1995, Ueno2005}  or non-perturbative functional renormalization group  \cite{ks_kpz_prep2024}.
At a numerical level, it allows one to observe the  KPZ universality emerging at large scales  for much smaller system sizes since the effective non-linearity $g_{\rm eff}$ is higher \cite{Ueno2005}.
 However, for the CGLE, the addition of noise is more subtle. 
The noisy CGLE, widely employed as a mean-field model for open quantum fluids \cite{Wouters2007, Carusotto2013}, exhibits a richer phase diagram than its deterministic version \cite{He2017, Vercesi2023}. In particular, it was shown that, at large noise, the compact nature of the phase becomes  crucial since topological defects (phase slips or space-time vortices in 1D) can be thermally activated by the noise, breaking down the analogy with growing interfaces. In tuning the noise amplitude $\sigma$, we are thus restrained to small values in order not to enter the vortex-turbulent phase of Ref.~\cite{He2017, Vercesi2023}.
We observe that the region of defect-free phase turbulence in the $(c_1, c_2)$ phase diagram shrinks as the noise amplitude is increased.
We could obtain such a regime only for small noise amplitude $\sigma \lesssim 0.1$ and by tuning $c_1, c_2$ closer to the BF instability line. 

The effect of the noise on the kinetic roughening properties of the phase is again encoded in the scale-dependent correlation time $\tau_{\alpha}(k)$, with $\alpha = C(\tau_{\alpha}, k)/C(k,0) \in (0, 1]$ (which is a slight generalization of the $\tau_{1/2}$ defined in Sec.~\ref{subsec:IB} corresponding to $\alpha=0.5$).
 We show in Fig.~\ref{fig:kpz_tau_noise_crossover} the behavior of $\bar{\tau} = \langle \tau_{\alpha} \rangle_{\alpha}$, where an average over $0.3 \leq \alpha \leq 0.6$ is performed in order to increase the statistics. First, we find that the IB regime with $z=1$  is always present at the  intermediate scales, robust to the addition of a small noise.
Let us now comment on the behavior at the large scales (small $k$ modes).
For the smallest value of the noise $\sigma = 0.01$, they follow the same scaling as for the deterministic case, i.e. we identify the EW regime with $\bar{\tau} \sim k^{-2}$. 
At  intermediate noise, here $\sigma = 0.05$, the EW regime is replaced by the KPZ one, with $\bar{\tau} \sim k^{-3/2}$. This corroborates the results found for the KS equation when adding noise.
 However, for stronger noise, here $\sigma=0.10$, the large scale behavior is affected by  the presence of defects, which eventually destroy the KPZ regime, since the phase can no longer be unwrapped. 
 In this regime we find  $\bar \tau\simeq$ constant as expected in presence of vortices. In fact, as shown in Refs.~\cite{He2017, Vercesi2023}, since defects are formed in randomly located space-time points, the phase trajectories are characterized by a finite homogeneous density of uncorrelated phase jumps. As a result, at length scales beyond the average vortex distance $l_v$, the phase dynamics can be interpreted as the result of a random deposition process, implying a scale-independent correlation time, i.e. $\tau(k \lesssim 2\pi / l_v) \sim $ constant.
 
 In order to fully characterize the KPZ regime for the intermediate noise value, we proceed as for the IB regime. We select the low modes for which $\bar{\tau} \sim k^{-3/2}$, and compute the full correlation $C(t, k)$ function in this window, shown in
 Fig.~\ref{fig:scaling_KPZ}. The level lines are now observed at constant $y=k^{3/2}t$ as expected for the KPZ regime. Moreover, a very good collapse is obtained, and the scaling function extracted from the numerical data compares accurately with the exact KPZ scaling function $f^{\rm KPZ}$ of Ref. \cite{Prhofer2004}.
 Thus, our results show that adding a noise, provided it is small enough to prevent the proliferation of defects, allows one to observe the KPZ scaling regime without having to resort to very large system sizes. This confirms that the phase turbulence of the CGLE equation belongs to the KPZ universality class. Moreover, the IB regime systematically appears at intermediate scales, and is an intrinsic feature of this system. Its origin can be traced to the necessary vanishing of the effective viscosity to crossover from a negative microscopic value to an effective positive value at large scales.

\section{\label{sec:Conclusions} Conclusions and perspectives}

We have studied the phase turbulence of the deterministic complex Ginzburg Landau equation in one spatial dimension, focusing on the statistical behavior of the large and intermediate wavelength modes. In this regime, the phase dynamics maps to the Kuramoto-Sivashinsky equation. This chaotic dynamics results in an effective noise, and generates
 an effective positive viscosity  at large scales. These elements, together with the intrinsic non-linearity of the phase dynamics, yield that the critical behavior  of the CGLE belongs to the 1D KPZ universality class.
 In our numerical simulations, we have first recovered the known results, namely we observe at large scales the EW scaling expected for the system size considered, smaller than the typical size necessary for the KPZ behavior to settle.

Focusing on the scales intermediate between the wavelength of the instability pattern and the onset of the EW (KPZ) behavior, we have evidenced the emergence of an additional, distinct scaling regime, characterized by the dynamical exponent $z=1$.
We have argued that this regime corresponds to the inviscid limit of the KPZ equation, which has recently been shown to be controlled by a genuine fixed point,  the inviscid Burgers one, of the KPZ equation in one dimension. Indeed, while the viscosity
crosses over from a negative value  at the microscopic scale in the KS equation to a positive value at the macroscopic scale in the effective KPZ equation, it has to vanish at some intermediate scale. This generates a  region of scales with vanishingly small viscosity,  and these scales are inherently controlled by the IB fixed point.
This explains the systematic appearance of the IB scaling $z=1$ in the CGLE, and KS equation.

We have also considered the noisy version of the CGLE, widely used in the context of driven-dissipative quantum fluids.
By focusing on the weakly unstable region of the parameter space, we have shown that  the KPZ scaling could be observed, enhanced by the presence of the noise, although the region of validity of the phase description shrinks due to noise-activated defects. This result allows us, on the one hand, to confirm that the phase turbulent regime of the CGLE belongs to the KPZ universality class.
On the other hand, it represents a yet unexplored regime of the noisy CGLE, for which the weakly unstable regime (phase turbulence) is found to be resilient to a small noise.  This opens up the route to potential applications to generic open systems in which the microscopic fluctuations are not negligible in the hydrodynamical description, such as exciton-polariton condensates.

\begin{acknowledgements}
We acknowledge Nicol\'as Wschebor and Marc Brachet for insightful discussions.
F.V. acknowledges the support from France 2030 ANR QuantForm-UGA and from the Laboratoire d’excellence LANEF in Grenoble (ANR-10-LABX-51-01).
A.M. acknowledges funding from the Quantum-SOPHA ANR Project ANR-21-CE47-0009-02. 
L.C. acknowledges support from Institut Universitaire de France (IUF).
\end{acknowledgements}

\appendix

\section{Stability analysis \label{app:linear_stability}}

The stability of the homogeneous rotating solution $\psi_0 = e^{-i c_2 t}$ is studied by considering the linearized evolution of plane waves on top of $\psi_0$. The dispersion is given by
\begin{align*}
    \omega^{\pm}(k) =& - i(1 + k^2) \; \pm \\
                    \pm& \sqrt{-(1+k^2)^2 + 2(1 + c_1c_2)k^2 + (1+c_1^2)k^4}\, .
\end{align*}
We define $\omega^{\pm}(k)= \epsilon^{\pm}(k) + i\gamma^{\pm}(k)$, where a positive growth rate $\gamma^{\pm}$ gives the instability condition. For small enough $k$, the square-root is purely imaginary. The rate $\gamma^{-}(k)$ is always negative, we thus focus on $\gamma^{+}(k)$ and expand it up to fourth order in $k$ to obtain
\begin{align*}
    \gamma^{+}(k) & = -(1+c_1c_2)k^2 - \frac{1}{2}c_1^2(1+c_2^2)k^4 + ... \\
    & = \nu k^2 + \tau k^4 + ...
\end{align*}
The instability onset is  controlled by the sign of $\nu = 1 + c_1c_2$. When $\nu<0$,  some low wavelength modes are unstable, approximately $0< k < k_0 = \sqrt{\nu/\tau}$.

\section{Phase equation \label{app:mapping_phase}}

In this Appendix, we detail the mapping from the CGLE to the phase equation.
The CGLE is given by
\begin{equation}
    i\partial_t \psi = i\psi + (c_2 - i)|\psi|^2 \psi - (c_1 - i) \partial_x^2 \psi.
    \label{eq:CGLE_c1c2_2}
\end{equation}
In  the amplitude-phase representation $\psi = \sqrt{\rho} \; e^{i\theta}$, it
 yields the two coupled equations for the amplitude and for the phase
\begin{equation}
    \begin{cases}
        \frac{\partial_t \rho}{2\rho} &= 1-\rho \; + \\
        & + \left\{ \frac{\partial_x^2 \rho}{2 \rho} - \frac{\left(\partial_x\rho \right)^2}{4\rho^2} - \left(\partial_x \theta\right)^2 \right\} - c_1 \left\{ \partial_x^2\theta + \frac{\partial_x\rho \partial_x\theta}{\rho} \right\}\\
        \partial_t \theta &= -c_2\rho  \; + \\
        & + \; c_1 \left\{ \frac{\partial_x^2 \rho}{2 \rho} - \frac{\left(\partial_x\rho \right)^2}{4\rho^2} - \left(\partial_x \theta\right)^2 \right\} - \left\{ \partial_x^2\theta + \frac{\partial_x\rho \partial_x\theta}{\rho} \right\} \,.
    \end{cases}\nonumber
    \label{eq:CGLE_rho_theta}
\end{equation}
We emphasize that small rigid ($\partial_x \, \cdot \,=0$) fluctuations of the amplitude around 1 have a relaxation time of order 1 (for this choice of units), while no such time-scale appears explicitly for the phase dynamics. This is a known consequence of the U(1) symmetry of the CGLE, which leads to assume that slow phase modulations dominate the fluctuations at the macroscopic scale.
The fast fluctuations of the amplitude around its stationary homogeneous value can thus be neglected, so that its modulations can be considered as enslaved to the slow phase dynamics.
By inserting the Ansatz $\rho = 1 + w(\theta, \partial_x^2 \theta, (\partial_x \theta)^2, ...)$ and substituting it into the amplitude equation, one obtains $w = \left( -c_1 \partial_x^2 \theta - \left(\partial_x \theta \right)^2 \right)$.
By inserting $w$ into the phase equation, considering linear terms up to order 4, we obtain:
\begin{equation}
    \partial_t \theta = (1 + c_1c_2) \partial_x^2 \theta - \frac{c_1^2}{2} \partial_x^4\theta +  (c_1 - c_2) \left( \partial_x \theta \right)^2 + ...
\end{equation}
which is the KS equation with $\nu = 1+c_1c_2$, $\tau = -\frac{c_1^2}{2}$ and $\lambda = 2(c_1 - c_2)$. The higher-order nonlinear terms, proportional to $\partial_x \theta \partial_x^3 \theta$ and $\partial_x^2\theta (\partial_x \theta)^2$, have been neglected. They are considered as unimportant in the long wavelength dynamics, while their effect is relevant for the short scales, when studying for instance the divergences which lead to the breakdown of the phase description \cite{Sakaguchi1990}.

We underline that the expression of $\tau$ obtained here is different from the one obtained from expanding the growth rate $\gamma^{+}(k)$ in perturbations in Appendix \ref{app:linear_stability}. The reason of the discrepancy is that $\gamma^{+}(k)$ can be considered as a phase-like branch only for small $k$. Unlike the pure phase diffusion $\nu k^2$, the order 4 term is affected since at increasing $k$ the off-diagonal perturbations, mixing phase and amplitude, become larger.

\providecommand{\noopsort}[1]{}\providecommand{\singleletter}[1]{#1}%

\end{document}